# New Model of Internet Topology Using $k$-shell Decomposition

Shai Carmi[1], Shlomo Havlin[1], Scott Kirkpatrick[2], Yuval Shavitt[3] and Eran Shir[3]

[1]*Minerva Center and Dept. of Physics, Bar-Ilan University, Ramat-Gan, Israel*
[2]*Benin School of Engineering and Computer Science, Hebrew University, Jerusalem, Israel*
[3]*Electrical Engineering Department, Tel-Aviv University, Israel*
(Dated: September 25, 2018)

We introduce and use $k$-shell decomposition to investigate the topology of the Internet at the AS level. Our analysis separates the Internet into three sub-components: (a) a nucleus which is a small ($\approx 100$ nodes) very well connected globally distributed subgraph; (b) a fractal sub-component that is able to connect the bulk of the Internet without congesting the nucleus, with self similar properties and critical exponents; and (c) dendrite-like structures, usually isolated nodes that are connected to the rest of the network through the nucleus only. This unique decomposition is robust, and provides insight into the underlying structure of the Internet and its functional consequences. Our approach is general and useful also when studying other complex networks.

The Internet has become a critical resource in our daily life. It still suffers from many inefficiencies, and as such has become a vibrant research subject [1, 2, 3]. Identifying the Internet's topology and its properties is a prerequisite to understanding its distributed, collaborative nature. Various tools from statistical physics, like scaling theory, percolation, and fractal analysis, have been applied to better understand the Internet and other complex networks [2, 3, 4, 5, 6, 7, 8, 9]. In particular, the surprising finding of the Internet's power-law degree distribution [10], has encouraged many scientists to use the degree (the number of immediate neighbors of a node) as an indicator of the importance and role of each node. However, using the degree as an indicator of function can be misleading both when looking at a single node and when looking at a distribution. For example, it has been shown [11] that topologies with a very different structure can have the same degree distribution.

Instead of node degree, we will use the "$k$-shell" decomposition to assign a shell index to each node in the internet. While node degrees can range from one or two up to several thousand, we find that this procedure splits the network up into 40-50 shells, the precise number depending upon the level of detail that our measurements provide. $k$-shell decomposition is an old technique in graph theory [12] and has been used as a visualization tool for studying long-scale networks such as the Internet [13]. It involves pruning the network down to those nodes with more than $k$ neighbors, just as has been studied in physics under the rubric of "bootstrap percolation," in the very different environment of regular lattices [14]. Other studies [15, 16, 17, 18], have developed the theory of some of the statistical properties of $k$-shells in random networks. We believe our application of this tool to discern the function of a node in such a graph is novel, as is the structure which we uncover.

In this Letter, we apply the $k$-shell decomposition to decompose the network into components with distinct functional roles. Surfacing the distinct role each component plays will demonstrate how this method helps us understand the large-scale function of a network as complex as the Internet. Also, it may reveal evolutionary processes that control the growth of the Internet [19].

Our Internet topology data-sets are among the first results of DIMES [20], a large-scale, distributed measurement effort to measure and track the evolution of the Internet. DIMES collects 3-6M measurements daily from a global network of over 10,000 software clients. We focus this paper on a set of measurements conducted from March through May, 2005. The results of DIMES' measurements can be analyzed to create several types of topologies, from the router level (where each node represents a single router on the Internet) to the Autonomous Systems (AS) topology (where each node is an entire sub-network, managed by a single organization, usually an Internet Service Provider (ISP)). This work will consider the high level (AS) topology, that results in a network containing $\approx 20000$ nodes.

Next we decompose the network into its $k$-shells. We start by removing all nodes with one connection only (with their links), until no more such nodes remain, and assign them to the 1-shell. In the same manner, we recursively remove all nodes with degree 2 (or less), creating the 2-shell. We continue, increasing $k$ until all nodes in the graph have been assigned to one of the shells. We name the highest shell index $k_{max}$. The $k$-core is defined as the union of all shells with indices larger or equal to $k$. The $k$-crust is defined as the union of all shells with indices smaller or equal to $k$.

We then divide the nodes of the network into 3 groups:

1. All nodes in the $k_{max}$-shell form the *nucleus*.

2. The rest of the nodes belong to the $(k_{max}-1)$-crust. The nodes that belong to the largest connected component of this crust form the *peer-connected component*.

3. The other nodes of this crust, which belong to smaller clusters, form the *isolated component*.

We show in Fig. 1 how the sizes of the two largest components in each $k$-crust vary with the crust index. A percolation transition is apparent, at $k = 6$. At this point the size of the second largest cluster and the average distance between nodes in the largest cluster are sharply peaked [21, 22]. This phase transition is similar to the transition found in [23] when removing the high degree nodes from a scale-free network. Above this point, the size of the largest cluster grows rapidly. At higher crusts it stabilizes, until it spans about 70% of the network at the $(k_{max} - 1)$-crust. When the nucleus is added, the network becomes completely connected.

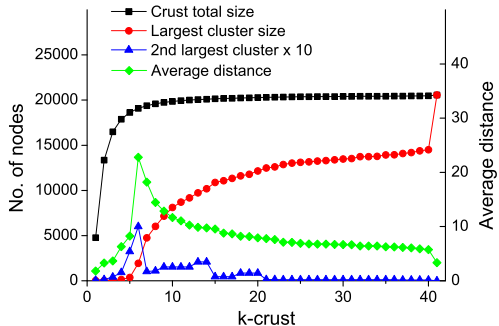

FIG. 1: For each $k$-crust, we plot the size of the crust (i.e. total number of nodes that belong to the crust), the size of the largest and 2nd largest connected components of the crusts, (2nd is magnified 10x to make it visible) and the average distance between nodes in the largest cluster of each crust.

This jump in connectivity and dramatic decrease in the distances observed at the $k_{max}$-shell justifies our definition of it as the nucleus. However, even in the absence of the nucleus, most of the network remains connected. This offers important opportunities for transport control over the Internet. For example, to avoid congestion in the nucleus, information can be sent using only the more peripheral nodes of the peer-connected component. Nevertheless, a significant fraction (roughly 30 per cent) of the network is not connected in the $(k_{max} - 1)$-crust. Those nodes, which form the isolated component, are either leaves or form small clusters, and can reach the rest of the network only through the nucleus. A schematic picture of the proposed decomposition is shown in Fig. (2). We call this a *Medusa* model, in part because of its apparent similarities with the "Jellyfish" model [24] proposed by Faloutsos et al., but our construction and the nature of its parts is different in each detail, as described below.

Identifying the nodes that form the heart of the Internet, the nucleus or "Tier One", is a problem that has been extensively investigated [24, 25, 26, 27]. For example, the nucleus might be defined as the set of all nodes with degree higher than some threshold. But this requires setting a free parameter, the degree threshold. Others [24] have defined it using a growing heuristic. Starting with an empty set, they add nodes to the nucleus in order of decreasing degree, retaining those for which the nucleus remains completely connected (a clique). Heuristics to build up a maximal clique are not robust. Moreover, node degree is an ambiguous indicator of importance. If we consider other reasonable orderings of the nodes [28], the resulting clique differs in over 25% of its constituent nodes. In contrast, our definition of the network's nucleus is unique, parameter-free, robust and easy to implement.

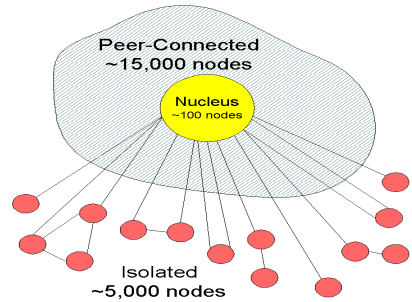

FIG. 2: A schematic plot of the suggested decomposition of the AS level Internet into three components. The structure of the plot resembles a jellyfish, thus we call it the *Medusa* model. (see below)

Analyzing the ASes that are found in our proposed nucleus, we find that the set which participates is very stable over time. We repeated the construction using data from three month intervals three and six months later than the data analyzed in this Letter, and found changes consisting of a few per cent of added sites, and one or two sites which moved from the nucleus to a $k$-shell immediately before it. The actual ASes involved include all major intercontinental carriers (about 10 nodes), plus carriers and Internet exchange points equally distributed among countries in North America, Europe, and the far East. The degree of nucleus sites ranged from over 2500 (ATT Worldnet) down to as few as 50 carefully-chosen neighbors, almost all within the nucleus (Google). The nucleus subgraph appears to be a dense random Erdős-Rényi graph, with diameter 2 and each node connected to roughly 70 percent of the other nucleus nodes.

An interesting question arises: does the size of the nucleus increase with the Internet size and how? Although we have seen a steady increase in the size of the AS graph during the course of the DIMES project, we cannot yet separate the actual growth of the Internet from the increase in our measurement sensitivity. Thus, we are led to investigate random ensembles of scale-free networks, with parameters (such as degree distribution) similar to the real Internet (Fig. 3(a)). Note, that the random graph calculation does not account well for the value of $k_{max}$ or the size of the nucleus, underestimating both by roughly an order of magnitude. However, the results suggest that the nucleus, as well as $k_{max}$, grows as a power of $N$. If in the limit of still larger random graphs, the nucleus is an Erdős-Rényi graph with a fixed fraction of bonds present, we would expect the two slopes to become the same, as can be seen in the figure.

The nodes in the peer-connected component can be connected without using the nucleus. This is an important property, since it enables communication without loading the nucleus. However, as we saw in Fig. 1, the nucleus provides shortcuts that decrease distances significantly. Several other interesting characteristics, such as scaling laws and fractal properties are found when analyzing the peer-connected component. For example, Fig. 3(b) shows the number of nodes of the peer-connected component coming from each shell. The decay follows a power-law with exponent 2.6 which can be derived by simple arguments [16]. When focusing on the $k$-crusts near the percolation threshold close to $k = 6$, we expect the connected part of the crust to show fractal properties [21]. In Fig. 4(a), we apply the box covering method, suggested by [5] to calculate the fractal dimension of networks, on the largest component of each crust. At the threshold the decay is a power law all along, with fractal dimension close to 2. It can be seen that for large $k$, the decay of the number of boxes needed to cover the network is exponential, indicating an infinite fractal dimension. A crossover length between a fractal and non-fractal regimes is seen when approaching the threshold ($k = 6$), as the decay of the number of boxes becomes a power-law with an exponential cutoff. Further support to the fractal picture is that we find that the degree distribution is invariant under box-renormalization [5], which indicates the property of self-similarity.

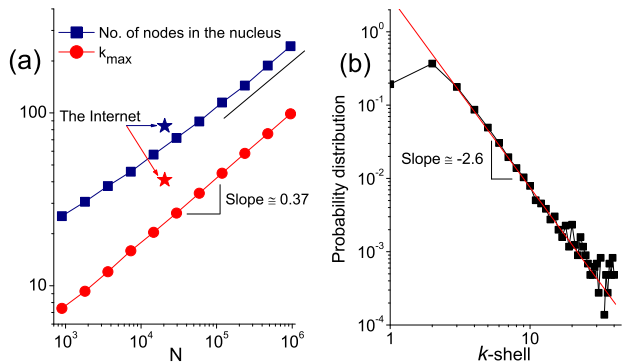

FIG. 3: (a) - The size of the nucleus and its $k$-shell index, as a function of $N$ is random scale-free networks. The values for the Internet are also indicated. (b) - The contribution of each shell to the peer-connected component.

The fractal dimension can be derived from arguments of percolation theory: At the threshold, almost all the high degree nodes are removed, such that the network becomes similar to a sparse random Erdős-Rényi network, as we have explicitly verified. Percolation in Erdős-Rényi random network is known to be equivalent to percolation in an infinite dimensional lattice, in which the fractal dimension of the largest component is 4, i.e. the mass of the largest cluster scales like $M \sim r^4 \sim l^2$. From percolation theory, we expect the probability distribution of (finite) cluster sizes to follow a power-law $p_s \sim s^{-\tau}$, with $\tau = \frac{5}{2}$ [21, 22]. Indeed, the $k$-crusts close to the percolation threshold show this behavior (see Fig. 4(b)).

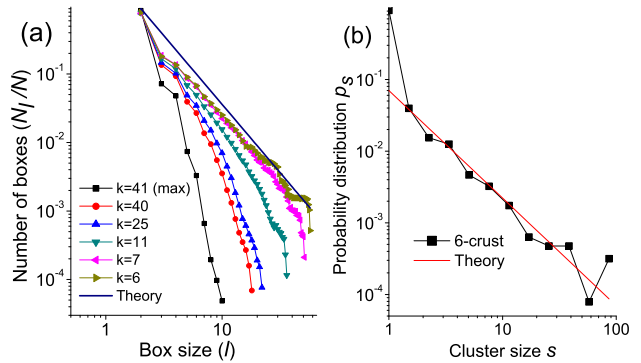

FIG. 4: (a) - For few selected crusts, we plot the number of boxes needed to cover the largest cluster of the crust as a function of the box sizes $l$. On a log-log scale, the slope of this curve is the fractal dimension of the network [5]. (b) - The probability distribution of the sizes of the finite clusters of the 6-crust. Percolation theory yields $p_s \sim s^{-5/2}$. (shown as a straight line) [21]

The isolated component includes nodes that are connected to the rest of the network through the nucleus only, and thus are usually not used in communication between the bulk of the network. Inspecting the nodes in this component, we find that they are either leaves (roughly one third of the nodes), or single nodes that connect with 2 or more links directly into the nucleus (60%), the rest (8%) form small clusters, with no more than 10 members. Schematically we can imagine these nodes as tendrils hanging from the center of the network, as do the tendrils of a jellyfish (See Fig. 2).

Such a jellyfish-like structure of the Internet at the AS level was first suggested in [24]. However, in [24], the nucleus is defined as a maximal clique, which is found to include only 20-25 ASes. The layers of the Jellyfish mantle are nodes a given number of hops from the nucleus and all leaves are considered tendrils. Defining the tendrils as all leaves does not capture the fact that a significant fraction of the network ($\approx 30\%$) form small clusters that are in a sense separated from the majority of the network, and that many leaves connect to the peer-connected part of the crust (our "mantle"). Moreover, since the Internet AS graph has a small diameter, the number of layers in the Jellyfish mantle is very small. Number of hops from the center is a much less sensitive measure than our k-shell index. Also, hop count obscures the fractal properties of the Internet which we observe. Thus, we claim that our model is a more detailed model for the topological structure of the Internet. Because of its Mediterranean origin, we call our decomposition a *Medusa* model.

Our proposed method of network analysis can be applied to other naturally occurring complex networks as well. Once decomposed, a careful examination of the network components - as the one carried out here for the Internet - can give insight into whether or not the network has the 'Medusa' properties. The studies with random long-scale networks reported above show that this structure should be common, but its quantitative details will differentiate models. Our parameters, for example, can serve to test of the validity of model Internet generators.

In [5] it was shown that realistic networks can be divided into two main groups : ones which posses fractal properties, and ones which do not. We show here that the Internet at the AS level, initially recognized as a non-fractal network (in [5] and here in Fig. 4), can be seen as composition of a fractal part and a nucleus. When the nucleus and the fractal part are joined, the nucleus provides the shortcuts that makes the Internet a small-world non-fractal network.

In summary, we have presented a method, based on the $k$-shell pruning, that is able to decompose a given network into three functionally distinct parts. We have applied our method to the latest and most accurate data on the Internet topology at the AS level, analyzing each part separately. We have gained much insight into the structure of the Internet, which is found to be Medusa-like, with a distinguished nucleus, the rest of the nodes either belong the largest fractal-like component, (which enables communication without congesting the nucleus), or to small clusters of nodes that are connected through the nucleus only. Our method is unique and robust and may assist in the analysis of other complex networks.

The DIMES measurements and our analysis are parts of the EVERGROW European integrated project No. 1935, funded within the Sixth Framework's Future and Emerging Technologies division. We also acknowledge support from the Israel Science Foundation, the Israel Internet Association, and the European NEST/PATHFINDER project DYSONET 012911. Discussions with Sorin Solomon, Avishalom Shalitt, and Alessandro Vespignani are also gratefully acknowledged. S.K. would like to thank ICSI (at U.C. Berkeley) for its hospitality during summer 2005, when parts of this work were carried out.

of their total number of links to nodes in the $k_{max}$ shell.